\documentclass[prd,amsmath,amssymb,longbibliography,superscriptaddress,twocolumn,nofootinbib,10pt]{revtex4-1}

\pdfoutput=1

\usepackage{graphicx}
\usepackage{dcolumn}
\usepackage{bm}
\usepackage{amssymb}
\usepackage{latexsym}
\usepackage{booktabs}
\usepackage{amsmath}
\usepackage{multirow}

\usepackage[colorlinks=true, linkcolor=blue, citecolor=blue]{hyperref}

\begin{document}

\title{Model-independent measurement of cosmic curvature with the latest $H(z)$ and SNe Ia data: A comprehensive investigation}

\author{Jing-Zhao Qi}
\affiliation{Key Laboratory of Cosmology and Astrophysics (Liaoning Province) \& Department of Physics, College of Sciences, Northeastern University, Shenyang 110819, China}
\author{Ping Meng}
\affiliation{Key Laboratory of Cosmology and Astrophysics (Liaoning Province) \& Department of Physics, College of Sciences, Northeastern University, Shenyang 110819, China}
\author{Jing-Fei Zhang}
\affiliation{Key Laboratory of Cosmology and Astrophysics (Liaoning Province) \& Department of Physics, College of Sciences, Northeastern University, Shenyang 110819, China}
\author{Xin Zhang}\thanks{Corresponding author.\\ zhangxin@mail.neu.edu.cn}
\affiliation{Key Laboratory of Cosmology and Astrophysics (Liaoning Province) \& Department of Physics, College of Sciences, Northeastern University, Shenyang 110819, China}
\affiliation{Key Laboratory of Data Analytics and Optimization for Smart Industry (Ministry of Education), Northeastern University, Shenyang 110819, China}
\affiliation{National Frontiers Science Center for Industrial Intelligence and Systems Optimization, Northeastern University, Shenyang 110819, China}

\begin{abstract}
{In the context of the discrepancies between the early and late universe, we emphasize the importance of independent measurements of the cosmic curvature in the late universe. We present an investigation of the model-independent measurement of the cosmic curvature parameter $\Omega_k$ in the late universe with the latest Hubble parameter $H(z)$ measurements and type Ia supernovae (SNe Ia) data. For that, we use two reconstruction methods, the Gaussian process (GP) and artificial neural network (ANN) methods, to achieve the distance construction from $H(z)$ data. Our analysis reveals that the GP method provides the most precise constraint on $\Omega_k$, with a constraint precision of $\xi(\Omega_k)=0.13$, surpassing recent estimations using similar methods. The GP method consistently indicates a preference for a flat universe at the 2$\sigma$ confidence level. Moreover, we find that the choice of reconstruction method influences the estimation of $\Omega_k$. The ANN reconstruction method exhibits higher sensitivity to the addition of BAO $H(z)$ data, resulting in comparable constraint precision to the GP method. A discrepancy exists between the best-fit values obtained by these two reconstruction methods, indicating their dependence on the reconstruction approach. However, we anticipate that with the improvement of sample size and precision of observational $H(z)$ data, the estimation of $\Omega_k$ using this approach will become more robust and reliable.}

\end{abstract}
\maketitle

\section{Introduction}
After nearly a century of cosmological research,  a standard cosmological model, the $\Lambda$ cold dark matter ($\Lambda$CDM) model with six base parameters, was established, whose validity has been verified by almost all current observational data \citep{SupernovaSearchTeam:1998fmf,SupernovaCosmologyProject:1998vns,WMAP:2003elm,SDSS:2003eyi,SDSS:2004wzw}, especially by the \textit{Planck}-satellite data with breathtaking precision \citep{Planck:2015fie,Planck:2018vyg}. However, as the precision of the observational data increases, some anomalies among different measurements of some key cosmological parameters are shaking our confidence in the standard cosmological model \citep{Riess:2019cxk,Planck:2018vyg,Qi:2020rmm,Qi:2022sxm,Cao:2021zpf,DiValentino:2019qzk,DiValentino:2020hov,Handley:2019tkm}. The most compelling ones appear to be the tension of the Hubble constant $H_0$ between  the value inferred from the \textit{Planck}-satellite data and the one inferred from the nearby type Ia supernovae (SNe Ia) data calibrated by the distance ladder, which has become a serious crisis for cosmology \citep{Riess:2020fzl,DiValentino:2021izs,Vagnozzi:2019ezj,Zhang:2019ylr,Xu:2020uws,Li:2013dha,Qi:2019zdk,Vattis:2019efj,Zhang:2014ifa,Guo:2018ans,Zhao:2017urm,Guo:2017qjt,Gao:2021xnk,Gao:2022ahg}. As far as the second anomaly is concerned, the $S_8$ parameter, a combination of the amplitude of matter density fluctuations and the matter density, is significantly lower in recent cosmic shear surveys than that expected according to the \textit{Placnk} data best fit $\Lambda$CDM model \citep{Heymans:2020gsg}.

In addition, recently, an enhanced lensing amplitude in cosmic microwave background (CMB) power spectra from \textit{Planck} 2018 data also presents another serious challenge to the standard cosmological model \citep{Planck:2018vyg,Handley:2019tkm,DiValentino:2019qzk}. This effect could be explained naturally by a closed universe. However, the prevailing and very successful inflationary theory predicts a flat universe, and the observations also support a flat universe. For instance, the combination of CMB power spectra data and baryon acoustic oscillation (BAO) data puts a stringent constraint on the cosmic curvature parameter $\Omega_k$ under $\Lambda$CDM, strongly supporting a flat universe \citep{WMAP:2003elm,Planck:2015fie,Planck:2018vyg,Stevens:2022evv}. As a result, most cosmological research has long assumed a flat universe. However, the discovery of a closed universe at more than 3.4$\sigma$ confidence level, preferred by the enhanced lensing amplitude in \textit{Planck} 2018 data, suggests that this assumption may not be taken for granted \citep{DiValentino:2019qzk,Handley:2019tkm}. What is worse is that the $H_0$ and $S_8$ tensions will be exacerbated when the possibility of a closed universe is considered, implying there may be even larger discordances hidden behind the assumption of a flat universe \citep{DiValentino:2019qzk,DiValentino:2020hov}. {For a comprehensive and detailed discussion on this topic, we recommend readers to refer to Refs. \citep{Efstathiou:2020wem,Vagnozzi:2020rcz,Vagnozzi:2020dfn,Gonzalez:2021ojp,Zuckerman:2021kgm,Akarsu:2021max,Glanville:2022xes,Bel:2022iuf,Stevens:2022evv,Favale:2023lnp,Cao:2021ldv,Zhai:2019nad,Ryan:2019uor,Park:2018tgj,Penton:2018wed,Ryan:2018aif,Yu:2017iju}}.

In fact, all of these crises arise from the measurement inconsistencies between the early and late universe, and point to the fact that the cracks have appeared in the standard $\Lambda$CDM model. It is time to reconfirm what we once knew for sure. In this paper, we aim to thoroughly investigate whether the spatial geometry of our universe is open, flat, or closed with a cosmological model-independent method. This is not only in response to the crisis mentioned above but also because the spatial curvature of the universe is a significant issue that is deeply relevant to many fundamental questions in modern cosmology, such as the evolution of the universe and the property of the dark energy. The spatial geometry of the universe is usually described by the cosmic curvature parameter $\Omega_k$, i.e., $\Omega_k>0$, $\Omega_k=0$ and $\Omega_k<0$ correspond to an open, flat, and closed universe, respectively. In order to tackle the crisis arising from the measurements of cosmic curvature, a necessary approach is to use a cosmological model-independent method to confirm the value of $\Omega_k$ in the late universe. Recently, great progress has been made in this regard.

So far, many cosmological model-independent methods have been proposed to determine the cosmic curvature parameter $\Omega_k$ \citep{Collett:2019hrr,Qi:2020rmm,Wang:2020dbt,Wang:2019vxv,Clarkson:2007bc,Clarkson:2007pz,Wang:2019yob,Xia:2016dgk,Cai:2015pia}, in which a popular and effective method is applying the distance sum rule in the combination of strong gravitational lensing (SGL) and SNe Ia observations to constrain  $\Omega_k$ \citep{Rasanen:2014mca}. Subsequently, this method has been fully implemented with the combination of SGL and other distance indicators, such as intermediate luminosity quasars and gravitational waves \citep{Wang:2022rvf,Qi:2020rmm,Qi:2022sxm,Cao:2021zpf}. However, while it is true that this method is independent of cosmological models, Qi et al. \citep{Qi:2018aio} found that it is dependent on the mass distribution model of lens galaxies and is also affected by the classification of SGL data according to the lens velocity dispersion. To obtain an unbiased and precise estimate  $\Omega_k$ in this way, it is crucial to accurately characterize the mass distribution of lens galaxies for each SGL sample, but this is still a long way off.

Another popular model-independent method to determine  $\Omega_k$ is originally proposed to test the homogeneous and isotropic Friedmann-Lema\^\i tre-Robertson-Walker (FLRW) metric \citep{Clarkson:2007pz}. This method is derived from the theoretical expression between cosmological distance and the Hubble parameter $H(z)$, in which the cosmic curvature is involved. In reverse, the cosmic curvature could be estimated under the assumption of the FLRW metric. Subsequently,  this estimation of $\Omega_k$ is implemented in several works in the light of new data and different statistical methods \citep{Wei:2016xti,Yu:2016gmd,Dhawan:2021mel}. In most of these works, the Gaussian process, a non-parametric reconstruction technique, is widely used to reconstruct a smooth curve of $H(z)$ so that the distance at any redshift can be calibrated. Recently, Wang et al. \citep{Wang:2019vxv,Wang:2020dbt} presented an alternative non-parametric approach based on the artificial neural network (ANN) for reconstructing a function from observational data, which also has been used in cosmological research, including the estimation of $\Omega_k$.

In view of the importance of the cosmic curvature and the advances in sample size and precision of observational data, this paper aims to thoroughly investigate the extent to which the cosmic curvature parameter can be constrained in the late-universe by using the available observational data and various reconstruction techniques. We will employ the latest Pantheon+ compilation of SNe Ia containing 1701 SNe Ia light curves and 60 $H(z)$ data obtained by two different observation methods to constrain the cosmic curvature parameter with two non-parametric approaches, GP and ANN.

\section{ Data and methodology}\label{2}
We dedicate this section to describing the methodology and two observational data sets used in this paper.
 
\subsection{Data: SNe Ia sample}
We use the SNe Ia Pantheon+ compilation containing 1701 light curves of 1550 unique in redshift range $0.001<z<2.26$ \citep{Brout:2022vxf}. Compared to the original Pantheon compilation \citep{Pan-STARRS1:2017jku}, the sample size of Pantheon+ compilation has greatly increased, and the treatments of systematic uncertainties in redshifts, peculiar velocities, photometric calibration, and intrinsic scatter model of SNe Ia also have been improved. It should be noted that not all the SNe Ia in the Pantheon are included in the Pantheon+ compilation.

In this paper, we make use of two different SNe Ia samples. Since the sensitivity of peculiar velocities is very large at low redshift ($z<0.008$) as shown in Fig.~4 of Ref.~\citep{Brout:2022vxf}, which may lead to biased results, we adopt the processing treatments used by Ref.~\citep{Brout:2022vxf}, namely, remove the data points in redshift range $z<0.01$. For convenience, we still call this data set as Pantheon+. In addition, the Pantheon+ dataset compiled by Ref.~\citep{Brout:2022vxf} also includes the recent Cepheid host distance anchors released by SH0ES (SNe, H0, for the equation of state of dark energy) Collaboration that facilitates constraints on both the standardized absolute magnitude of the SNe Ia $M$ and $H_0$. Here, we also use this SNe Ia data set and call it as Pantheon+\&SH0ES.

{While it is true that, in principle, the combination of $H(z)$ data with SNe Ia data can provide constraints on $\Omega_k$, the inclusion of the SH0ES Cepheid calibration plays a crucial role in our analysis for several reasons. Firstly, the determination of $H_0$ is highly sensitive to the constraint on $\Omega_k$ in our analysis. By combining the SH0ES Cepheid calibration, we benefit from an independent and remarkably precise measurement of $H_0$, which significantly enhances the accuracy of determining $\Omega_k$. Secondly, the SH0ES Cepheid data allow for the calibration of the absolute magnitude of SNe Ia, providing a valuable anchor for cosmological distance measurements. This calibration is instrumental in achieving more reliable distance estimates. Finally, the synergy achieved by combining multiple independent probes, such as the SH0ES Cepheid data, $H(z)$, and SNe Ia, facilitates a more robust and comprehensive determination of $\Omega_k$, minimizing potential biases and yielding more confident results.}

For each SN Ia, the observed distance module is given by 
\begin{eqnarray}
\mu_{\mathrm{SN}}=m_B-M_B,\label{con:E2.1}
\end{eqnarray}
where $m_B$ is the observed magnitude in the rest-frame \textit{B}-band. The theoretical  distance modulus $\mu_{\mathrm{th}}$ is defined as
\begin{eqnarray}
\mu_{\mathrm{th}}=5\log_{10}\left[\frac{D_L(z)}{\mathrm{Mpc}}\right]+25,\label{con:E2.2}
\end{eqnarray}
where ${D_L(z)}$ is the luminosity distance associated with the cosmological parameters. Constraining cosmological parameters is implemented by minimizing the $\chi^2$ function:
\begin{equation} \label{Eq:chi2}
-2 \ln (\mathcal{L})=\chi^2=\Delta D^T \mathbf{C}_{\text {stat }+\text { syst }}^{-1} \Delta D,
\end{equation}
where $D$ is the SNe Ia distance-modulus residuals computed as
\begin{equation}
\Delta D = \mu_{\text{SN}}-\mu_{\text{th}},
\end{equation}
and $\mathbf{C}_{\text {stat }+\text { syst }}$ is the covariance matrix including both statistical and systematic errors, which could be found in the website\footnote{https://github.com/PantheonPlusSH0ES/DataRelease}, as well as the SH0ES Cepheid host-distance covariance matrix.

\subsection{Data: Hubble parameter measurements}

The Hubble parameter $H(z)$ describes the expansion rate of the universe, and its observation is an important probe for exploring dark energy and the evolution of the universe. In general, there are two ways to measure $H(z)$. One is obtained by calculating the differential ages of galaxies, which is called cosmic chronometer (CC). We denote this $H(z)$ data obtained by this method as CC $H(z)$ \citep{Zhang:2012mp,Stern:2009ep,Moresco:2012jh,Moresco:2016mzx,Ratsimbazafy:2017vga,Moresco:2015cya}. Another is inferred from the baryon acoustic oscillation (BAO) peak in the galaxy's power spectrum. For convenience, we call this $H(z)$ as BAO $H(z)$ \citep{Gaztanaga:2008xz,Oka:2013cba,BOSS:2016zkm,Chuang:2012qt,BOSS:2016wmc,BOSS:2013rlg,Blake:2012pj,Zhao:2018gvb,Busca:2012bu,Bautista:2017zgn,BOSS:2014hwf,BOSS:2013igd}. We have compiled the latest 32 CC $H(z)$ data points in Table \ref{T1} and 28 BAO $H(z)$ data points in Table \ref{T2}. In this paper, we preferentially use CC $H(z)$ data to constrain $\Omega_k$, and then employ the total $H(z)$ data (CC $H(z)$ + BAO $H(z)$). {Considering the importance of properly accounting for the correlations between data points, in our analysis, we use the publicly available CC covariance tool \footnote{https://gitlab.com/mmoresco/CCcovariance} to estimate the covariance matrix for the CC data \citep{Moresco:2012jh,Moresco:2015cya,Moresco:2016mzx,Moresco:2020fbm}. This allows us to appropriately incorporate the correlations and uncertainties associated with the CC measurements into our analysis. Furthermore, we have reanalyzed the results by considering the updated CC covariance and have found consistent results, reinforcing the robustness of our findings. The covariance matrix of the CC data is computed as}
\begin{equation}
\textbf{Cov}_{ij}=\textbf{Cov}_{ij}^{\text{stat}}+\textbf{Cov}_{ij}^{\text{sys}},
\end{equation}
{where $\textbf{Cov}_{ij}^{\text{stat}}$ is the statistical errors. The systematic uncertainties $\textbf{Cov}_{ij}^{\text{sys}}$ encompass various effects associated with the determination of physical properties of galaxies, such as stellar metallicity and potential contamination from a young component. These effects are uncorrelated for objects at different redshifts.  For a more comprehensive understanding of the origin and modeling of systematic errors in the CC data, we refer readers to Ref. \citep{Moresco:2020fbm}.}

\begin{table}
\caption{The CC Hubble parameter $H(z)$ measurements and their errors $\sigma_{H(z)}$ at redshift $z$ obtained from the differential age method.}\label{T1}
\centering
\begin{tabular}{ccccc}
\hline
 Index &  $z$  &  $H(z)$ [Mpc] & $\sigma_{H(z)}$ [Mpc]  & Reference \\
\hline
1 &   0.07   &   $69.0$  & $19.6$  & \citep{Zhang:2012mp}    \\
2 &   0.1    &   $69.0$   & $12.0$    & \citep{Stern:2009ep}   \\
3 &   0.12   &   $68.6$  & $26.2$ & \citep{Zhang:2012mp}    \\
4 &   0.17   &   $83.0$ & $8.0$  & \citep{Stern:2009ep}    \\
5 &   0.1797   &   $81.0$ & $5.0$  & \citep{Moresco:2012jh}   \\
6 &   0.1993   &   $81.0$ & $6.0$  & \citep{Moresco:2012jh}   \\
7 &   0.2    &   $72.9$ & $29.6$  & \citep{Zhang:2012mp}    \\
8 &   0.27 &   $77.0$ & $14.0$  & \citep{Stern:2009ep}    \\
9 &   0.28    &   $88.8$  & $36.6$    & \citep{Zhang:2012mp}   \\
10 &   0.3519  &   $88.0$ & $16.0$ & \citep{Moresco:2012jh}    \\
11&   0.3802  &   $89.2$ & $14.1$  & \citep{Moresco:2016mzx}    \\
12&   0.4  &   $95.0$& $17.0$  & \citep{Stern:2009ep}   \\
13&   0.4004   &   $82.8$ & $10.6$  & \citep{Moresco:2016mzx}   \\
14&   0.4247    &   $93.7$ & $11.7$  & \citep{Moresco:2016mzx} \\
15&   0.4293   &   $91.8$ & $5.3$  & \citep{Moresco:2016mzx}    \\
16&   0.4497 &   $99.7$  & $13.4$    & \citep{Moresco:2016mzx}   \\
17 &   0.47   &   $89.0$& $49.65$  & \citep{Ratsimbazafy:2017vga}    \\
18&   0.4783    &   $80.9$& $9.0$  & \citep{Moresco:2016mzx}    \\
19&   0.48   &   $97.0$ & $60.0$  & \citep{Stern:2009ep}   \\
20&   0.5929   &   $110.0$ & $15.0$  & \citep{Moresco:2012jh}   \\
21&   0.6797 &   $98.0$ & $10.0$  & \citep{Moresco:2012jh} \\
22&   0.7812 &   $88.0$ & $11.0$  & \citep{Moresco:2012jh}    \\
23&   0.8754 &   $124.0$  & $17.0$    & \citep{Moresco:2012jh}   \\
24&   0.88 &   $90.0$ & $40.0$ & \citep{Stern:2009ep}    \\
25&   0.9    &   $117.0$ & $23.0$  & \citep{Stern:2009ep}    \\
26&   1.037   &   $113.0$ & $15.0$ & \citep{Moresco:2012jh}   \\
27&   1.3   &   $168.0$ & $17.0$  & \citep{Stern:2009ep}   \\
28&   1.363 &   $160.0$ & $33.6$  &\citep{Moresco:2015cya} \\
29&   1.43  &   $177.0$ &$18.0$  & \citep{Stern:2009ep}    \\
30&   1.53   &   $140.0$ &  $14.0$ &  \cite{Stern:2009ep}   \\
31&   1.75  &   $202.0$ & $40.0$  & \citep{Stern:2009ep}   \\
32&   1.965    &   $186.5$ & $50.4$  & \citep{Moresco:2015cya} \\
\hline
\end{tabular}
\end{table}

\begin{table}
\caption{The BAO Hubble parameter measurements $H(z)$ and their errors $\sigma_{H(z)}$ at redshift $z$ obtained from the radial BAO method.}\label{T2}
\centering
\begin{tabular}{ccccc}
\hline
 Index &  $z$  &  $H(z)$ [Mpc] & $\sigma_{H(z)}$ [Mpc] & Reference\\
\hline
1 &   0.24   &   $79.69$  & $2.99$ &\citep{Gaztanaga:2008xz}\\
2 &   0.3    &   $81.7$   & $6.22$ &\citep{Oka:2013cba} \\
3 &   0.31   &   $78.17$  & $4.74$ &\citep{BOSS:2016zkm} \\
4 &   0.34   &   $83.80$   & $3.66$ &\citep{Gaztanaga:2008xz}\\
5 &   0.35   &   $82.70$   & $8.40$  &\citep{Chuang:2012qt} \\
6 &   0.36   &   $79.93$  & $3.39$ &\citep{BOSS:2016zkm} \\
7 &   0.38   &   $81.50$   & $1.90$  &\citep{BOSS:2016wmc} \\
8 &   0.40   &   $82.04$  & $2.03$ &\citep{BOSS:2016zkm} \\
9 &   0.43   &   $86.45$  & $3.68$ &\citep{Gaztanaga:2008xz}\\
10&   0.44   &   $84.81$  & $1.83$ &\citep{BOSS:2016zkm} \\
11&   0.48   &   $87.79$  & $2.03$ &\citep{BOSS:2016zkm} \\
12&   0.51   &   $90.40$   & $1.90$  &\citep{BOSS:2016wmc} \\
13&   0.52   &   $94.35$  & $2.65$ &\citep{BOSS:2016zkm} \\
14&   0.56   &   $93.33$  & $2.32$ &\citep{BOSS:2016zkm} \\
15&   0.57   &   $96.80$   & $3.40$  &\citep{BOSS:2013rlg} \\
16&   0.59   &   $98.48$  & $3.19$ &\citep{BOSS:2016zkm} \\
17&   0.6    &   $87.90$   & $6.10$  &\citep{Blake:2012pj} \\
18&   0.61   &   $97.30$   & $2.10$  &\citep{BOSS:2016wmc} \\
19&   0.64   &   $98.82$  & $2.99$ &\citep{BOSS:2016zkm} \\
20&   0.73   &   $97.30$   & $7.00$    &\citep{Blake:2012pj} \\
21&   0.978  &   $113.72$ & $14.63$&\citep{Zhao:2018gvb} \\
22&   1.23   &   $131.44$ & $12.42$&\citep{Zhao:2018gvb} \\
23&   1.526  &   $148.11$ & $12.71$&\citep{Zhao:2018gvb} \\
24&   1.944  &   $172.63$ & $14.79$&\citep{Zhao:2018gvb} \\
25&   2.3    &   $224.00$    & $8.00$    &\citep{Busca:2012bu} \\
26&   2.33   &   $224.00$    & $8.00$    &\citep{Bautista:2017zgn}\\
27&   2.34   &   $222.00$    & $7.00$    &\citep{BOSS:2014hwf} \\
28&   2.36   &   $226.00$    & $8.00$    &\citep{BOSS:2013igd} \\
\hline
\end{tabular}
\end{table}

\begin{figure*}
\includegraphics[scale=0.4]{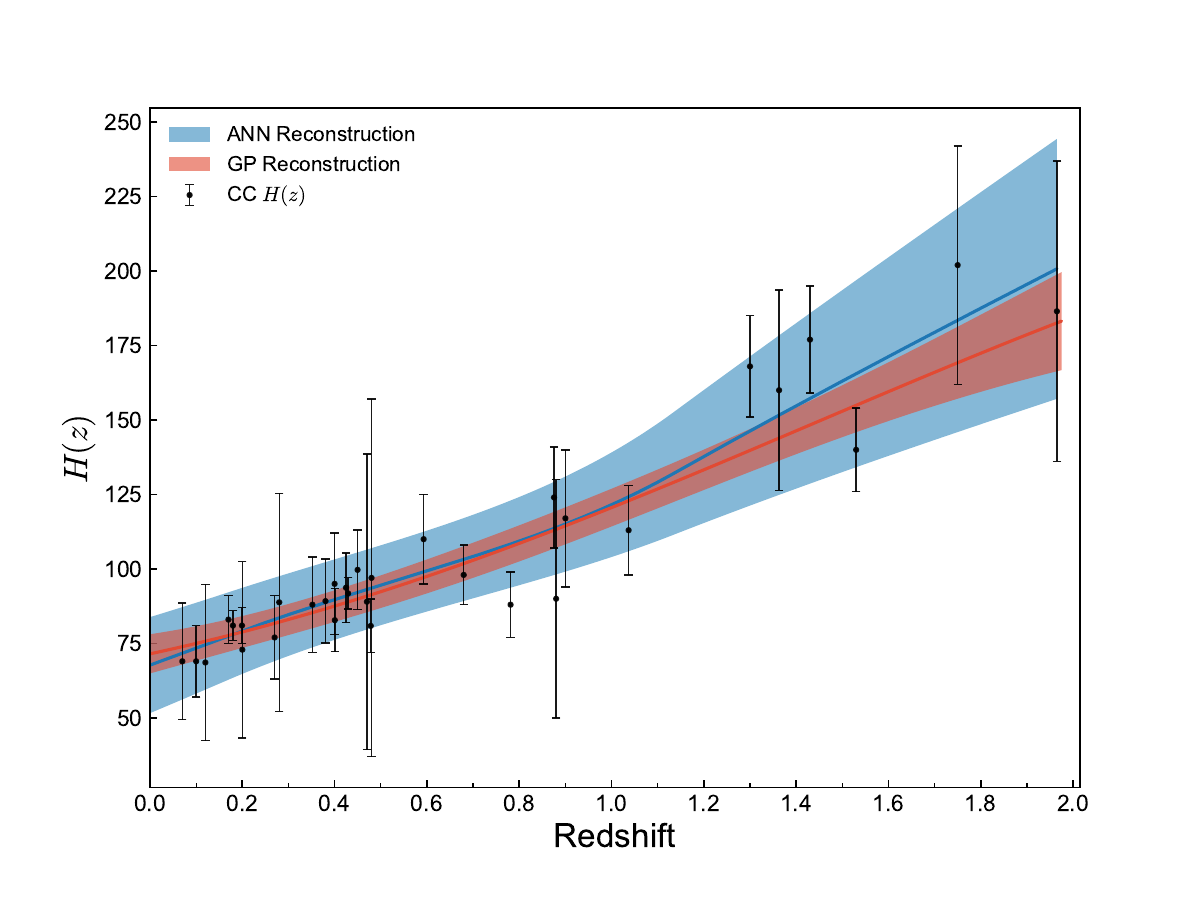}
\includegraphics[scale=0.4]{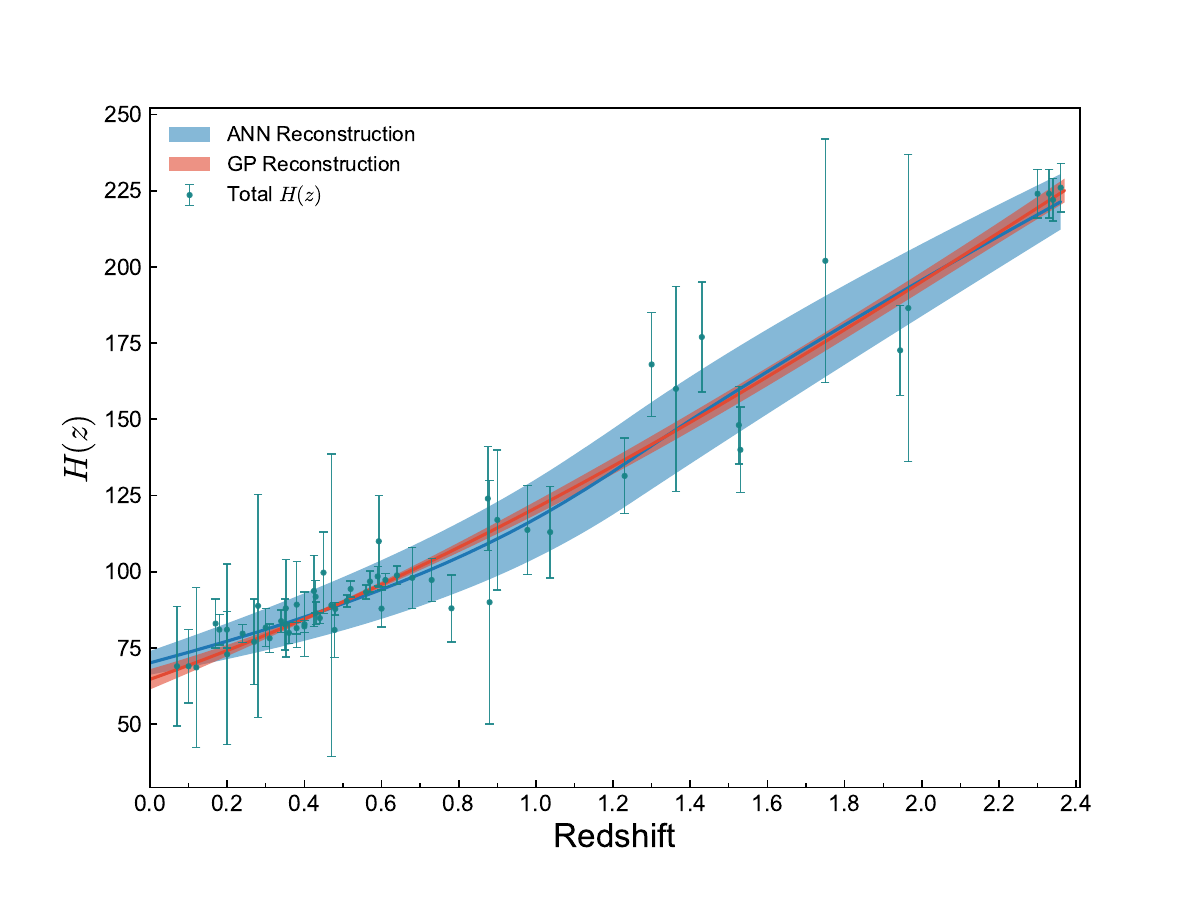}
\caption{Left: The reconstructions of $H(z)$ from CC $H(z)$ data by using GP (red region) and ANN (blue region). The shaded region and the solid line denote the 1$\sigma$ confidence level errors and the best-fit values of reconstruction, respectively. The black points with error bars represent the observed CC $H(z)$ data. Right: Same as the left panel but for the Total $H(z)$ data.} 
\label{fig:reconstruction}
\end{figure*}

\subsection{Reconstruction method: Gaussian process}
There is an integral between the Hubble parameter and the luminosity distance. In order to calibrate the distance using the $H(z)$ data, it is necessary to reconstruct a smooth curve of $H(z)$ with a non-parametric reconstruction technique firstly. Here, we briefly introduce the GP method that allows us to reconstruct a function from data straightforwardly without any parametric assumption. We adopt the \texttt{GaPP} Python code widely used in cosmology to implement the GP method \citep{Seikel:2012uu, Seikel:2012cs,Zhang:2018gjb,Cai:2019bdh,Wang:2020dbt,Seikel:2013fda,Benisty:2022psx,Briffa:2020qli,Bernardo:2021cxi,Escamilla-Rivera:2021rbe}. In this process, it is assumed that the value of the reconstructed function $f(z)$ evaluated at two different points $z$ and $\tilde{z}$ are connected by a covariance function $k(z,\tilde{z})$, and it only depends on two hyperparameters $\sigma_f$ and $\ell$. Although there are various and effective forms of the covariance function, according to the analysis in Seikel \& Clarkson \cite{Seikel:2013fda}, the squared exponential form with the Mat\'{e}rn $(\nu=9/2)$ covariance function can lead to more reliable results than all others. So we take it here, and its expression is
\begin{eqnarray}
k(z,\tilde{z})&=&\sigma_f^2\exp(-\frac{3|z-\tilde{z}|}{\ell})\nonumber\\
&\times &(1+\frac{3|z-\tilde{z}|}{\ell}+\frac{27(z-\tilde{z})^2}{7\ell^2}\nonumber\\
&+&\frac{18|z-\tilde{z}|^3}{7\ell^3}+\frac{27(z-\tilde{z})^4}{35\ell^4}).\label{7}
\end{eqnarray}
{Here, the hyperparameter $\ell$ represents the characteristic length scale, indicating the distance over which significant changes occur in the function $f(z)$. The hyperparameter $\sigma_f$ represents the typical change or variation in the observed data. The values of two hyperparameters are optimized by the GP itself via the observational data. It is important to note that the optimization of these two hyperparameters is performed independently of the fitting process for the cosmological parameters.}
The reconstructed functions of $H(z)$ for the two cases, CC $H(z)$ and Total $H(z)$, are shown in Fig. \ref{fig:reconstruction}.

\subsection{Reconstruction method: artificial neural network}
Here, we use the ANN method based on \texttt{REFANN} \citep{Wang:2019vxv} Python code to reconstruct a function of $H(z)$ from data, which also has been widely used in cosmology \citep{Dialektopoulos:2021wde,Benisty:2022psx}. The ANN method, completely driven by data, allows us to reconstruct a function from any kind of data without assuming a parametrization of the function. The optimal ANN model of reconstructing functions we used is the same as that selected by Wang et al. \citep{Wang:2019vxv}, which has one hidden layer with 4096 neurons total. The reconstructed functions of $H(z)$ from the ANN are also shown in Fig.~\ref{fig:reconstruction}. 

We can see that the confidence region reconstructed from the ANN is larger than that reconstructed from the GP method, which may be due to the basic logic and nature of these two techniques. {One potential explanation for the observed discrepancy is the variance in the underlying assumptions and modeling approaches of GP and ANN. The GP method focuses on reconstructing a smooth function based on the covariance between data points, prioritizing the overall structure and correlations in the data. On the other hand, ANN approximates the underlying function using interconnected artificial neurons, allowing it to capture complex non-linear relationships. These inherent differences in modeling techniques can lead to variations in how the methods handle noise, outliers, and subtle features in the data, resulting in divergent reconstructions for $H(z)$. Additionally, the training and optimization processes employed by GP and ANN can contribute to the observed discrepancy. The selection of hyperparameters, such as the kernel function in GP or the network architecture in ANN, can significantly impact the models' flexibility and generalization capabilities. Variations in the hyperparameter selection and training strategies may introduce sensitivities and biases that influence the inferred values of $H(z)$. Furthermore, it is important to acknowledge that GP and ANN have distinct strengths and limitations. GP excels at capturing uncertainties and estimating smooth functions, while ANN is effective in modeling complex non-linear relationships. These inherent differences in methodology can contribute to the observed discrepancies in the reconstructed values of $H(z)$.}

In addition, the reconstructed values of Hubble constant ($H(z=0)$) by these two methods are also different, which is sensitive to the constraint on $\Omega_k$ as we will see later.

\subsection{Methodology for estimation of $\Omega_k$}
In the framework of the FLRW metric, the comoving distance $D_C(z)$ is defined as
\begin{eqnarray}
D_C(z)=c\int_{0}^{z}\frac{d z^{\prime}}{H\left(z^{\prime}\right)}\label{con:E2.4},
\end{eqnarray}
where $c$ is the speed of light. With a reconstructed smooth function of $H(z)$, a smooth function of $D_C(z)$ could be calculated by integrating the function $H(z)$, and its confidence region also could be obtained by integrating the error of $H(z)$. Furthermore, the luminosity distance $D_L$ could be obtained by $D_C$ via
\begin{equation}
\frac{D_L}{(1+z)}= \begin{cases}\frac{c}{H_0}\frac{1}{\sqrt{\Omega_k}} \sinh \left[\sqrt{\Omega_k} D_C \frac{H_0}{c}\right] & \Omega_k>0, \\ D_C & \Omega_k=0, \\ \frac{c}{H_0}\frac{1}{\sqrt{\left|\Omega_K\right|}} \sin \left[\sqrt{\mid \Omega_k} \mid D_C \frac{H_0}{c}\right] & \Omega_k<0. \end{cases}
\end{equation}
Note that in this calculation, the value of $H_0$ is adopted from the reconstructed value of $H(z=0)$. The uncertainty of $D_L$ could be obtained by
\begin{eqnarray}
\sigma_{D_L}= \begin{cases}(1+z) \cosh \left[\sqrt{\left|\Omega_{k}\right|} D_C \frac{H_0}{c}\right] \sigma_{D_C} & \text {for} \quad\Omega_{k}>0, \\ (1+z) \sigma_{D_C} & \text {for} \quad\Omega_{k}=0, \\ (1+z) \cos \left[\sqrt{\left|\Omega_{k}\right|} D_C\frac{H_0}{c}\right] \sigma_{D_C} & \text {for} \quad\Omega_{k}<0.\end{cases}\label{con:E2.7}
\end{eqnarray}
The distance modulus reconstructed from $H(z)$ data $\mu_H$ can be further obtained by Eq.~(\ref{con:E2.2}). Finally, the cosmic curvature parameter $\Omega_k$ could be estimated by minimizing the $\chi^2$ function of Eq.~(\ref{Eq:chi2}). Here, the uncertainty of reconstructed distance modulus $\sigma_{\mu_{H}}$ should be added to the covariance matrix as a systematic error via 
\begin{equation}
(\mathbf{C}_{\text{stat}})_{ii}=(\mathbf{C}_{\text{stat}}^{\text{SN}})_{ii}+\sigma_{\mu_{H},i}^2.
\end{equation}
We constrain the cosmological parameters using the \texttt{emcee} Python module based on the Markov Chain Monte Carlo analysis \citep{Foreman-Mackey:2012any}. There are two free parameters, $\Omega_k$ and the SNe Ia absolute
magnitude $M_B$.

\section{Results and discussions}\label{3}
\begin{figure*}
\includegraphics[scale=0.45]{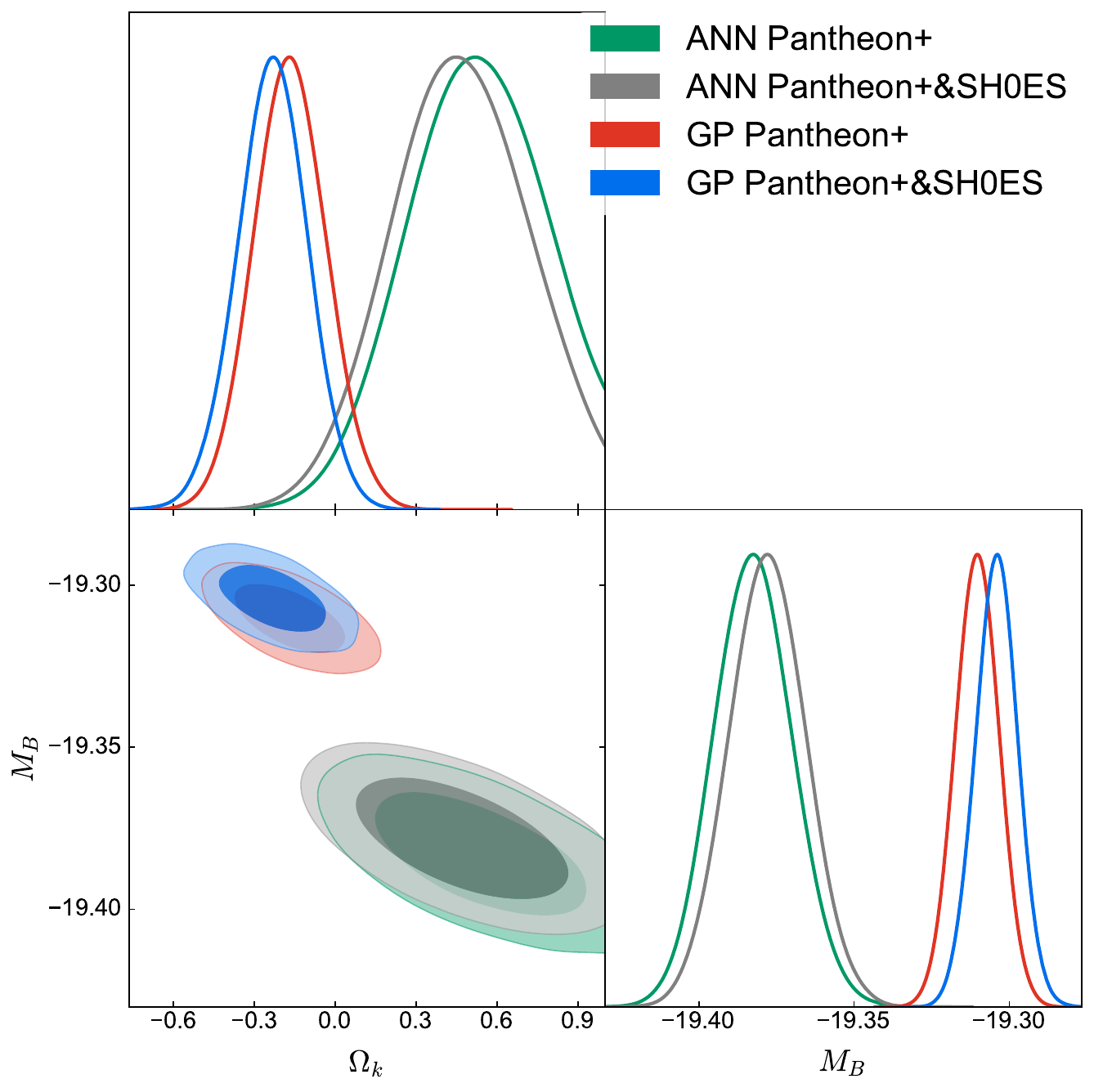}
\caption{The constraints with the 1$\sigma$ and 2$\sigma$ confidence level on the cosmic curvature $\Omega_k$ and the SNe Ia absolute magnitude $M_B$  from two types of SNe Ia data set (Pantheon+ and Pantheon+\&SH0ES) and with two reconstruction methods (GP and ANN), respectively, in the case of CC $H(z)$ data.} 
\label{CC_results}
\end{figure*}

\begin{figure*}
\includegraphics[scale=0.45]{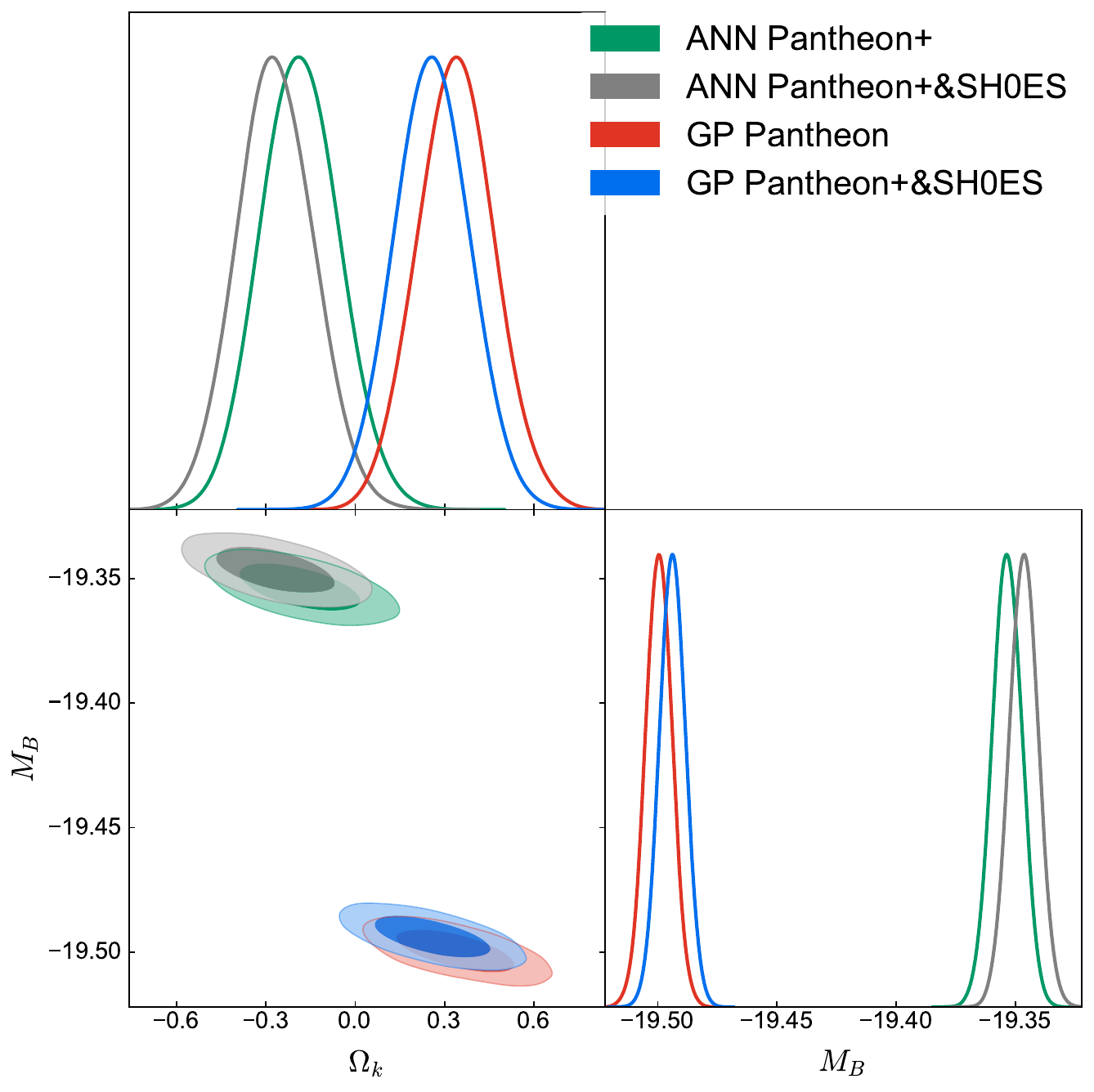}
\caption{Same as Fig. \ref{CC_results}, but for Total $H(z)$ data.} 
\label{total_results}
\end{figure*}

\begin{table*}  
\renewcommand\arraystretch{1.5}
\caption{The constraint results of $\Omega_k$ and $M_B$ with 1$\sigma$ confidence level for various data sets and reconstruction methods.}\label{Table}
\center
\begin{tabular}{|c|c|c|c|c|c|}
\hline
$H(z)$ data type & Parameters & GP Pantheon+ & GP Pantheon+\&SH0ES & ANN Pantheon+ & ANN Pantheon+\&SH0ES \\ 
\hline
\multirow{2}{*}{ CC $H(z)$ }& $\Omega_k$ & $ -0.17\pm 0.13$ & $-0.23\pm 0.13$ & $0.51^{+0.27}_{-0.24}$ & $0.46\pm 0.25$ \\
 & $M_B$ & $-19.310\pm 0.007$ & $-19.304\pm 0.007$ &$-19.382\pm 0.013$ & $-19.378\pm 0.012$\\
\hline
\multirow{2}{*}{ Total $H(z)$ }& $\Omega_k$ &$0.34\pm 0.13$ & $0.26\pm 0.13$ & $-0.18\pm 0.13$ & $-0.27\pm0.13$ \\
 & $M_B$ & $-19.500\pm 0.006$ & $-19.494\pm 0.005$ & $-19.353\pm 0.006$  & $-19.3465\pm 0.006$  \\
\hline
\end{tabular}
\end{table*}

Here, we combine two types of $H(z)$ data, two types of SNe Ia data, and two reconstruction methods to make a thorough investigation of the cosmic curvature. All the constraint contours of $\Omega_k$ and $M_B$ are shown in Figs.~\ref{CC_results}--\ref{total_results} and the best-fit values with 1 $\sigma$ confidence level  are listed in Table~\ref{Table}.

{In Fig.~\ref{CC_results}, we present the constraints on $\Omega_k$ and $M_B$ obtained from the CC $H(z)$ data in different scenarios. It is evident that there are differences between the contours derived from the two reconstruction methods. Specifically, concerning the GP method, the values of $M_B$ constrained from both the Pantheon+ and Pantheon+\&SH0ES data sets are almost identical, which means the addition of SH0ES data is not helpful to the constraint on $M_B$. As for $\Omega_k$, the estimations from Pantheon+ and Pantheon+\&SH0ES both favor a closed universe, while remaining consistent with a flat universe at the 2$\sigma$ confidence level. Moreover, we find an intriguing finding that the inclusion of SH0ES data does not appear to significantly impact the precision of the $\Omega_k$ constraint.}

Regarding the ANN method, we find that the uncertainties of the two parameters (i.e., $\Omega_k$ and $M_B$) obtained from this approach are larger than those derived from the GP method. This discrepancy can be traced back to the left panel of Fig.~\ref{fig:reconstruction}, where we notice that the confidence region of $H(z)$ reconstructed by the ANN is notably broader compared to that reconstructed using the GP method. Additionally, as mentioned previously, the reconstructed values of the Hubble constant ($H(z=0)$) from these two methods are also different, which subsequently influences the constraint on $M_B$ as evident from the distinct contours observed in Fig.~\ref{CC_results} for the GP and ANN methods. Concerning the estimation of $\Omega_k$, we find that not only do the uncertainties become larger when compared to the results obtained from the GP method, but the best-fit values also tend to favor an open universe, while remaining consistent with a flat universe within the 2$\sigma$ confidence level. Additionally, the addition of SH0ES data does not appear to significantly affect the constraint precision of $\Omega_k$ in the context of the ANN method.

Now, let's focus on the Total $H(z)$ data, which has a sample size nearly twice as large as the CC $H(z)$ data. In Fig.~\ref{total_results}, we present the 1D and 2D marginalized probability distributions of $\Omega_k$ and $M_B$ obtained from this dataset.

{For the GP method, regarding the constraints on $\Omega_k$, we find a shift in the best-fit values compared with CC $H(z)$ data, now leaning towards a positive value, indicating support for an open universe, but the estimate from Pantheon+\&SH0ES is still consistent with a flat universe at 2$\sigma$ confidence level. This change demonstrates that the addition of the BAO $H(z)$ observational data influences the estimation of $\Omega_k$ using this approach. Despite the Total $H(z)$ data having nearly twice the sample size of the CC $H(z)$ data, we find that the constraint precision of $\Omega_k$ is not notably improved compared to that derived from the CC $H(z)$ data. }

In the case of the ANN reconstruction method, we find a substantial improvement in the constraints on both $\Omega_k$ and $M_B$ when using the Total $H(z)$ data compared to the results obtained from the CC $H(z)$ data. The constraints on $\Omega_k$ and $M_B$ have improved by approximately twice, indicating that the ANN method is more sensitive to the addition of the BAO $H(z)$ data. This sensitivity allows for better constraints on the cosmological parameters when incorporating the larger sample size provided by the Total $H(z)$ data. Regarding $\Omega_k$, both of the best-fit values from the two SNe Ia data sets favor a closed universe, while remaining consistent with a flat universe within the 2$\sigma$ confidence level.

\section{Conclusion}\label{4}

Currently, the measurement inconsistencies between the early and late universe, such as the tensions in the Hubble constant, the $S_8$ parameter, and the cosmic curvature parameter, have raised questions about the validity of the standard cosmological model, i.e. the $\Lambda$CDM model. In this paper, we highlight the importance of determining the cosmic curvature parameter $\Omega_k$ and aim to make a thorough investigation for the model-independent measurement of $\Omega_k$ in the late universe with the observational data and statistical tools available to us. Therefore, we consider two types of $H(z)$ data sets (CC $H(z)$ data and Total data), two types of SNe Ia data sets (Pantheon+ and Pantheon+\&SH0ES), and two reconstruction methods (GP method and ANN method).

{The GP method has yielded the most precise constraint on $\Omega_k$, with a constraint precision of $\xi(\Omega_k)=0.13$ for any combination of data, surpassing the recent measurements of $\Omega_k$ using similar methods \citep{Dhawan:2021mel,Wei:2016xti,Yu:2016gmd,Wang:2020dbt}. Overall, the estimations obtained through the GP method consistently support a flat universe at the 2$\sigma$ confidence level.}

{It is worth noting that the estimation of $\Omega_k$ in this study is influenced by the choice of reconstruction method. The ANN reconstruction method exhibits higher sensitivity to the addition of $H(z)$ data. By combining the BAO $H(z)$ data, the constraint precision based on the ANN method becomes comparable to that obtained using the GP method. However, a discrepancy exists between the best-fit values obtained by these two reconstruction methods, indicating a dependence on the reconstruction approach. As a consequence, the method employed in this study to evaluate $\Omega_k$ may be less robust due to its sensitivity to the reconstruction method. Nevertheless, we expect that with the improvement of sample size and precision of observational $H(z)$ data, the estimation of $\Omega_k$ using this approach will become more robust and reliable.}

\section*{Acknowledgments}
This work was supported by the National SKA Program of China (Grants Nos. 2022SKA0110200 and 2022SKA0110203),  and the National Natural Science Foundation of China (Grants Nos. 12205039, 11975072, 11835009, and 11875102).
\bibliography{curvature_hzsn}

\end{document}